# A DISCRETE ANALYSIS OF METAL-V BELT DRIVE

**A. KARAM** and **D. PLAY**

Université Libanaise
Faculté des sciences II
FANAR-MATEN B.P. 90656
Lebanon

## Abstract

The metal-V belt drive includes a large number of parts which interact between them to transmit power from the input to the output pulleys. A compression belt composed of a great number of struts is maintained by a tension flat belt. Power is them shared into the two belts that moves generally in opposite directions. Due to the particular geometry of the elements and to the great number of parts, a numerical approach achieves the global equilibrium of the mechanism from the elementary part equilibrium. Sliding arc on each pulley can be thus defined both for the compression and tension belts. Finally, power sharing can be calculated as differential motion between the belts, is defined. The first part of the paper will present the different steps of the quasi-static mechanical analysis and their numerical implementations. Load distributions, speed profiles and sliding angle values will be discussed. The second part of the paper will deal to a systematic use of the computer software. Speed ratio, transmitted torque, strut geometry and friction coefficients effect will be analysed with the output parameter variations. Finally, the effect pulley deformable flanges will be discussed.

## I - Introduction

The metal **V-belt** of **VAN-DOORNE** type [1], is used in **C.V.T.** mounted on cars. The pulleys are kept at a fixed axis distance and are subjected to axial forces which can be varied in order to adapt the radius of the belt wrapped on the pulleys so as to obtain different speed ratios.

The belt (Fig. 1 a-b) consists of one set of thin flexible steel bands (or flat belt) on which rides a large number of small steel blocks (struts), whose lateral surfaces are in contact with the flanges of the pulleys. During the motion, a strut on the driving pulley is carried forward by the pulley via the interfacial friction between the pulley and the strut. Thus, a strut is pressed on the preceeding strut, generating a compressive force between them. This compressive force increases as the struts travel from the leading edge to the trailling edge of the driving contacting arc. On the driven pulley, this compressive force decreases until the end of the contact arc. Power is then transmitted from the driving pulley to the driven pulley through the linear compressed struts zone (upper part of the figure, Fig.1). Further, the flat belt do not only serve as a track constraints for the travelling struts, but also works as a part of the power transmission. The tight side of the flat belt is generally located at the lower linear zone (Fig.1).

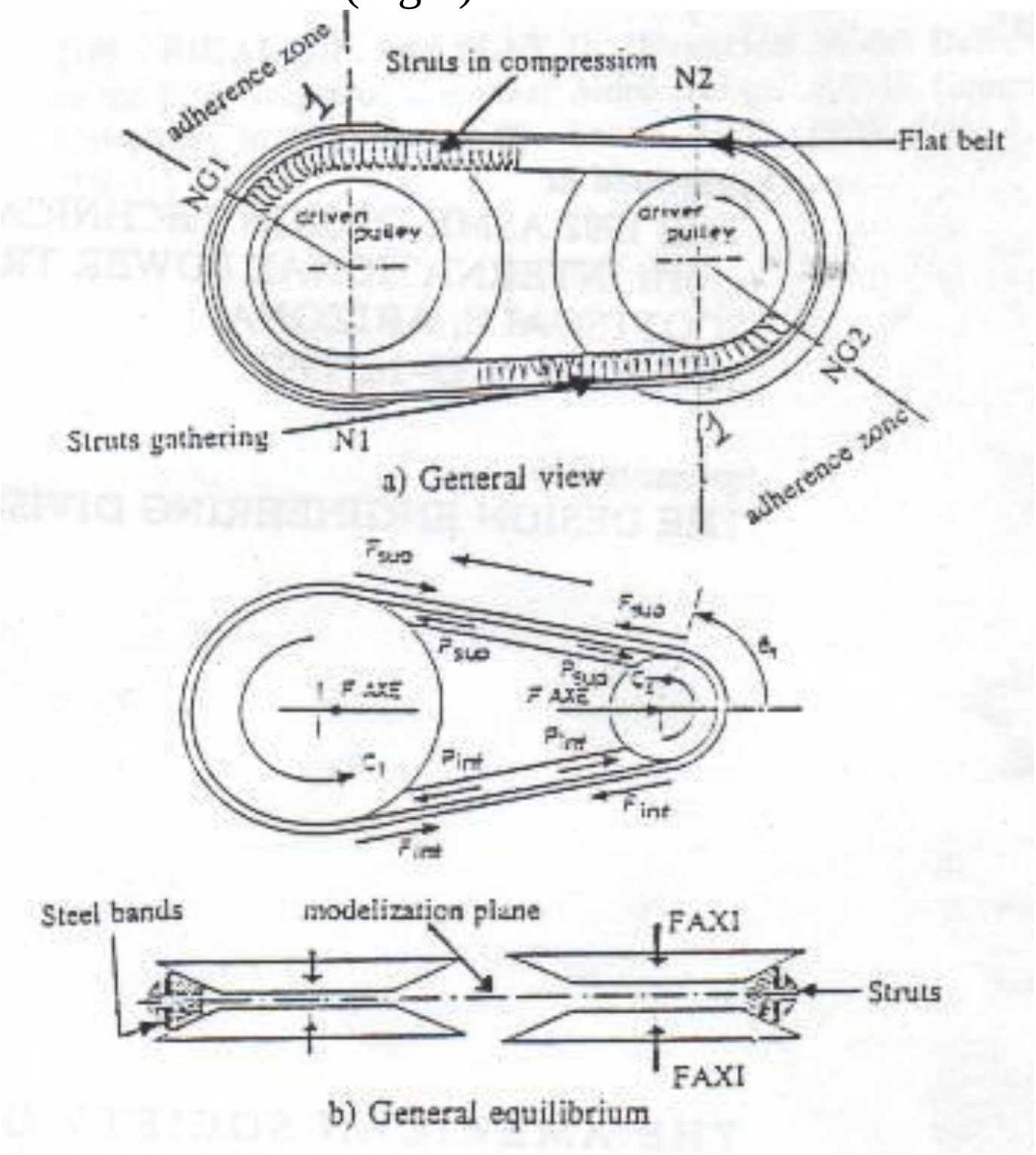

Fig. 1: Geometry and Modelization of C.V.T. transmission

A fine knowledge of the behaviour of each mechanical part is obviously necessary to insure the life and the reliability of the system. As the experimental tests become more and more expensive, numerical models permit to give data for design considerations. Thus final experimental tests serve only for last minor modifications. Moreover, optimization of the car performances passes through the interactions between the mechanical system and the micro-processored actuators. For exemple, the axial force applied on the moveable flanges of the pulleys must be adapted in order to insure the transmitted torque conditions with a maximum life of the elements. Thus, the knowledge of mechanical governing parameters becomes necessary to design efficient actuators.

The litterature on metal V-Belt behaviour appeared during the last decade [3-4-5]. All the papers are based on the wellknown approach of the V-belt behaviour proposed by Gerbert [6]. Their purposes were to establish models for this highly decretized V-belt. Various hypothesis have been made in order to get a first understanding of the behaviour. The purpose of this paper is to make a generalization of a model proposed in 1987 by the author [5] in order to obtain more details on the mechanical parameter variations. New geometries are also included and consequently the equations are redevelopped. This paper will present:

- a theoretical mechanical analysis with the classical hypothesis of general mechanics,
- a calculating numerical process and the resulting computing architecture,
- results and discussion.

## II - Theoretical analysis and resolution

The modelling of a mechanical system is complicated as the number of parts is large. The interaction between parts are also amplified if the stiffness variations gives changes of the initial geometry. On the other hand, into a metal V-belt drive, the set of parts is moving and the relative motions depend on the working conditions and the whole mechanical equilibrium. It is then necessary to describe simultaneously the equilibrium of all parts and the kinematic conditions. Due to the large number of parts, the resolution of the equations must be numerical. As the initial conditions are not fixed, the numerical resolution is iterative. In this first paragraph, the steps of calculations are summarized in a such way to facilitate the numerical resolution. The details of the modelization are given elsewhere [7].

### II-1 Hypotheses

The following hypotheses have been made in order to modelize the system:

- the steel struts are always in compression that means an initial tension is considered when the metal V-belt is at rest,
- the two sets of bands are identified as one flat belt which holds the struts in a symetrical manner, the analysis is thus bi-dimensionnal,
- the flat belt is very rigid and its length is thus, considered constant during the motion. The strut are seeing more deformable due to their geometry.
- the flanges of the pulleys will be considered rigid during the analysis. Note that a discussion on the effect of the deformable flanges will be made at the end of the paper.

The model takes also into consideration the following assumptions:

- the coefficients of friction of flat belt/struts and struts/pulleys is considered through the Coulomb model at this level of the mechanical analysis of the behaviour,
- the radial and tangential displacements of struts occur relatively to the flat belt and the flanges of the pulleys,
- full rotations of struts take place at their entry in the contacting arc, without futher variation of angular position until the end of the contact,
- the strains of struts follow a classical bulk material law.

After these hypotheses, the theorical analysis is detailled in three paragraphs:- Strut equilibrium - flat belt equilibrium - global equilibrium.

## II-2 <u>Strut equilibrium</u>

During the study of strut equilibrium, the following simplified hypotheses are established:

- a strut stays in a radial position, AiBi constitutes linear cord, AiA'i and BiB'i are supposed to be transversal contacts lines between the struts [(i+1),i] and [i,i-1],
- the action between the pulley and a strut is characterized by a squew applied on the contact pulley strut surface at the Mpi point (fig. 2-a),
- the inertia forces, operating only on the contacting arcs of the pulleys are radial.

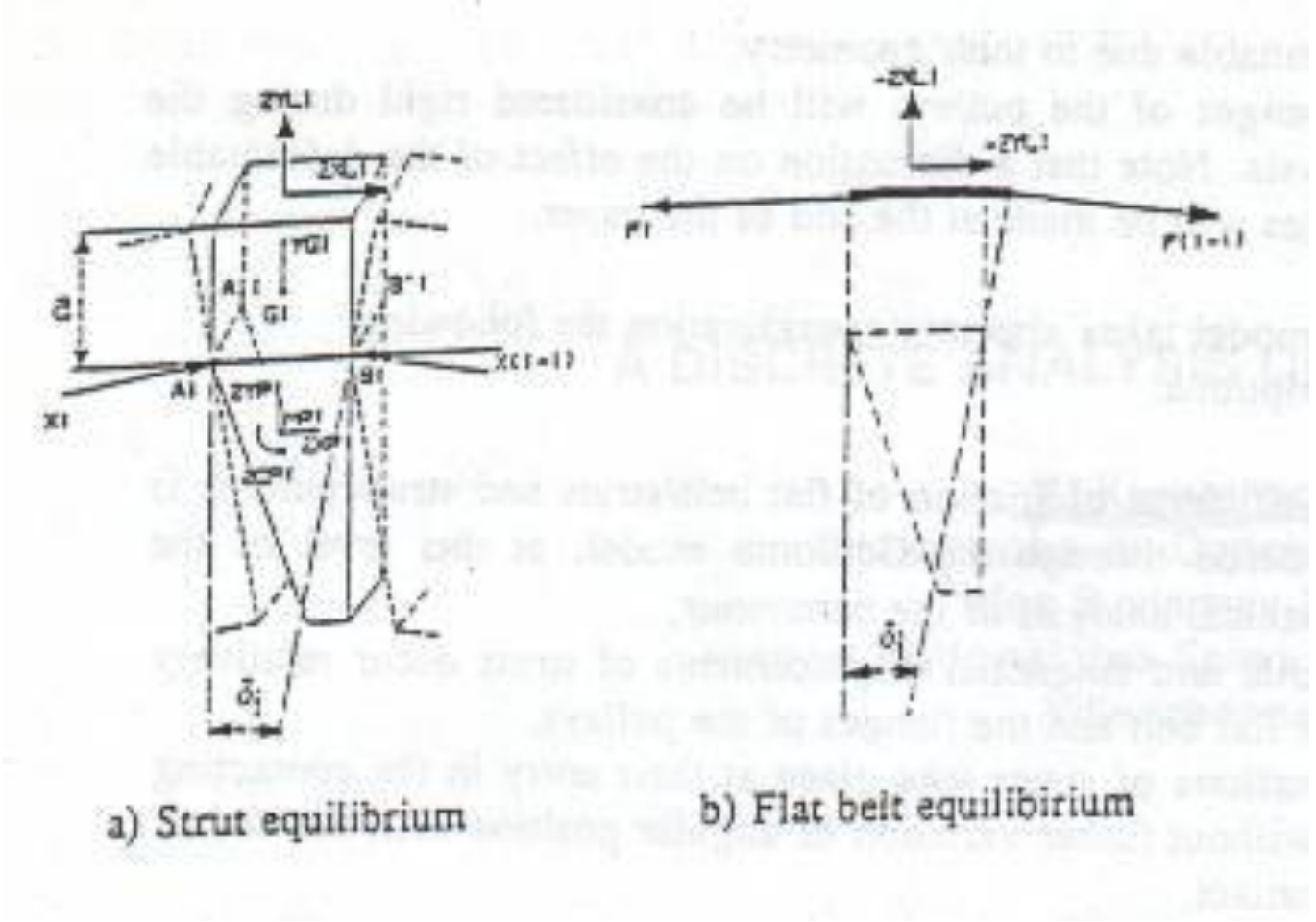

Fig. 2: Free body definition

By application of general mechanical rules, the following equations are obtained:

$$-X_{i-1} + X_i \cos(\delta_j) + 2\, XL_i + 2\, Xp_i = 0$$

$$X_i \sin(\delta_j) + 2YL_i + 2YP_i + YG_i = 0$$

$$-2yp_i\, XP_i + 2xp_i\, YP_i + x_G\, YG_i + 2xl_i\, YL_i + R(\cos(\delta j)-1)X_{i-1} + 2CPi = 0$$

where $(xp_i,yp_i)$ is a coordinate of $MP_i$, xG is an abscissa of G, $xl_i$ = l/2, torque $CP_i$ is applied on a strut. Thus, this third equation is used only for the determination of $CP_i$.

## II-3 **Belt equilibrium**

The internal forces considered into the flat belt are longitudinal traction forces. Bending strengh and cordal effect are neglected. The equilibrium of a flat belt element which spreads out on a strut (fig. 2-b) gives the following equations:

$$F_{i-1} \cos(\delta_j) - F_i - 2\, XL_i \quad = 0$$
$$F_{i-1} \sin(\delta_j) + 2\, YL_i \quad = 0$$

The forces $XL_i$ and $YL_i$ applicated by the strut on the flat belt element are related to the Coulomb's model and to the direction of the relative motion between the two mechanical elements.

$XL_i = \mu L\, \varepsilon L_i\, YL_i$ where $\varepsilon L_i$ is a coefficient equal to $\pm 1$ whose the sign is defined by the relative motion.
By substitution, the preceeding equations give:

$$F_i = AL_i\, F_{i-1} \text{ where } AL_i = (\cos \delta_i + \mu L\, \varepsilon L_i\, \sin \delta_i)$$

If $N_1$, $N_2$ are respectively the number of struts into the contact arc 1 and 2, the global equilibrium of the flat belt is expressed by:

$$F_{N2,2} = Fsup = F_{1,1}$$
$$F_{N1,1} = Finf = F_{1,2}$$

For j = 1,2 (j=1 : driven pulley, j=2 : driving pulley):

$F_{i+1,j} = AL_{i,j}\, F_{i,j}$. Then writing all the forces:

$$Finf = \prod_{i=1}^{N1} AL_{i,1}\ Fsup; \qquad Fsup = \prod_{i=1}^{N2} AL_{i,2}\ Finf$$

The result of the substitution is:

$$\prod_{i=1}^{N1} AL_{i,1} \prod_{i=1}^{N2} AL_{i,2} = 1$$

This equation will be used later to determinate the average velocity VL of the flat belt.

## II-4 <u>Struts/pulleys Interaction</u>

The contact between pulley/strut generates tangential force $TP_i$ and a normal force $NP_i$ which are related by the Coulomb's model: $|TP_i| < \mu p\, |NP_i|$.

TPi is situed into the contact plane (strut-pulley). It makes an angle $\gamma_{P_i}$ (called sliding angle) with the tangential direction (Fig. 3-a). The components are related by:

$$YP_i = NP_i \sin\phi + TP_i \cos\phi \sin(\gamma_{P_i})$$
$$XP_i = TP_i \cos\gamma_{P_i})$$
$$ZP_i = NP_i \cos\phi - TP_i \sin\phi \sin(\gamma_{P_i})$$
$$RP_i = XP_i\, tg(\gamma_{P_i})$$

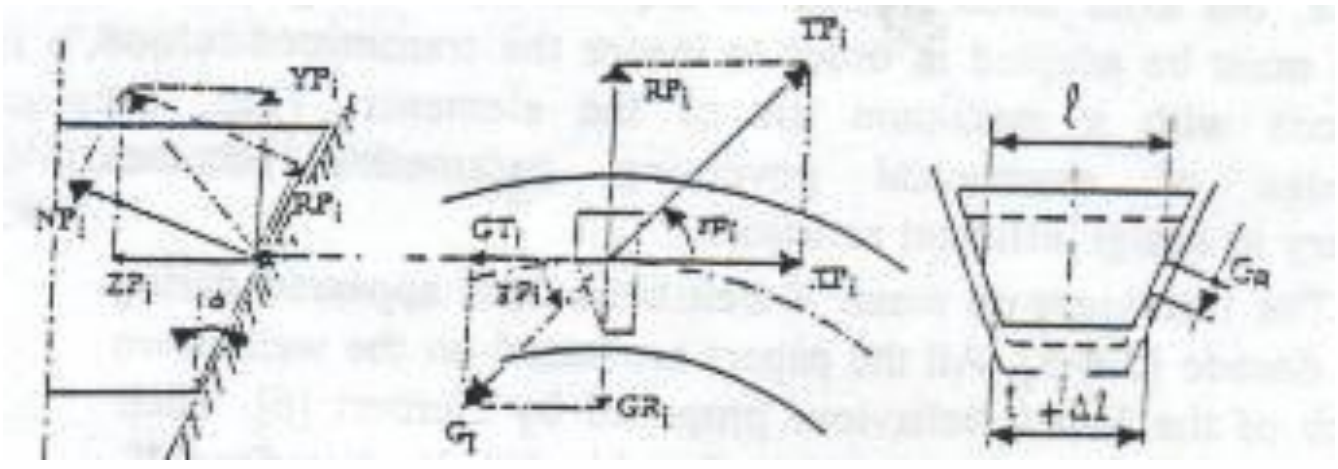

a) Slipping of struts between the conic flanges of the pulleys.

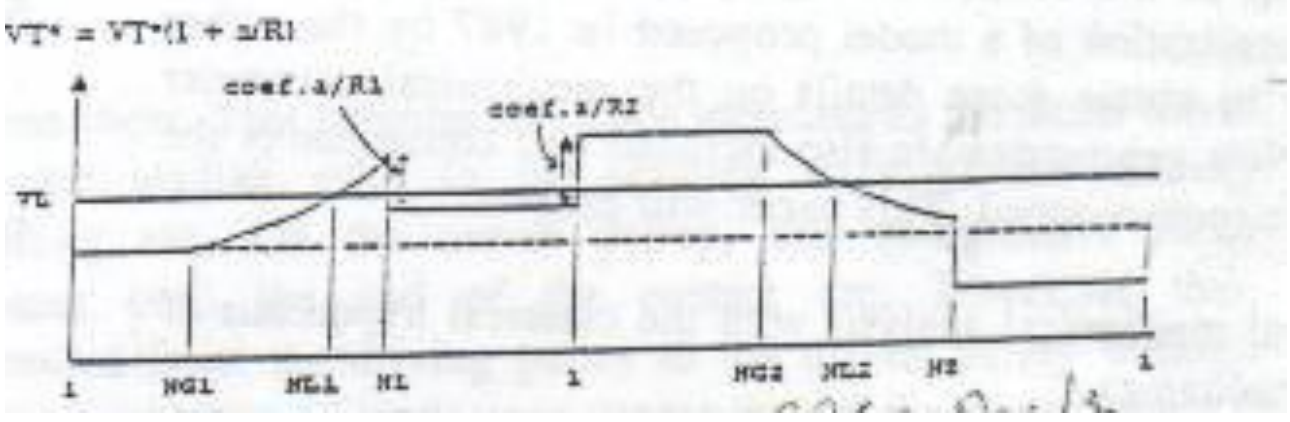

b) Velocity profile of struts and flat belt.

fig. 3: Modelization of kinematic conditions

Each contact arc is subdivised in two zones:

- an adherence zone, where the actions between the struts $X_i$ and the actions of pulleys on these struts stay constant:

$$X_i = X_{i-1}$$
$$ZP_i = ZP_{i-1}$$

- a sliding zone, where the forces decrease on the driven pulley and increase on the driving pulley.

$$XP_i = \mu p_i \, YP_i$$

with $\mu p_i = [\mu p \cos(\gamma_{p_i})]/[\sin\phi + \mu p \sin(\gamma_{p_i}) \cos\phi]$

The sliding angle $\gamma_{p_i}$ is defined by the deformations of a strut from a position (i) to an other position (i+1):

$\gamma_{p_i}$ = Arctg[$-GR_i/GT_i$] where $GR_i$ and $GT_i$ are repectively the radial and tangential motion of the strut. Note that GT and GR are calculated from a global experimental nonlinear relation between load and displacement (Fig. 3-a).

## II-5 Strut Velocities

The struts keep a constant velocity along the adherence zones of the contact arcs. Along the sliding zones, the velocity varies in an opposite manner to the actions $X_i$. Eurther, we assure that GR and GT are independant parameters.

$VT_i = l_i/(\Delta t) = VT_{i-1} + (\Delta li)/(\Delta t)$ so:

$VT_i = ([1 - (\Delta l_{i-1})/l]/[1 - (\Delta l_i)/l])\ VTi\text{-}1$

The velocity profile (Fig. 3-b) can be deduced. Note that at each entry and exit of the belt circular zones on the pulleys, a discontinuity apears because of the strut contact location change. Thus, a parameter $a/R_i$ must be introduced.

## II-6 Flat belt velocity

The flat belt is supposed to be rigid longitudinally comparing to the strut deformations. A constant average velocity VL is thus considered. A presupposed known velocity profile is considered (that means a known struts velocity) in order to calculate this velocity (Fig. 3-b). The two struts NL1 and NL2 are considered to have the same velocity which is the looked for one. With the global equilibrium equation of the flat belt and the passing over all contact arcs, determination of VL becomes possible.

## II-7 Global Equilibrium of C.V.T.

The two linear strands and two contact arcs constitute the four zones along the belt. The flat belt equilibrium in each of these zones can be described by knowing the forces imposed on the extremites of the considered zones (fig. 1-b). By isolating the two parts (driving and driven) of the C.V.T., the following equations are written:

$$FAXE = (Fsup + Finf - Psup - Pinf)\sin(\theta_1)$$

$$C_2 = (R_2 + GR_2)\,Psup - R_2\,Pinf - (R_2 + a + GR_2)Fsup + (R_2 + a)Finf$$

$$C_1 = (R_1 + GR_1)Pinf - R_1\,Psup - (R_1 + a + GR_1)Finf + (R_1 + a)Fsup$$

## II-8 Power and efficency calculation

These calculations are made when all the system equilibriums are established. If the torque on the two pulleys and the rotation velocity of each pulley are knwon, the power calculation is obvious:

$PW_2 = C_2\ W_2$ and $PW_1 = C_1\ W_1$

The efficency is given by the following expression:

$\mu = -PW_1/PW_2$

In the upper and lower strands, the forces PSUP, PINF, FSUP and FINF are transmitted from the driving pulley to the driven pulley. Then, in these strands, no power loss is considered. The global loss of power is obtained by PW1 + PW2 with PW1 < 0.

## II-9 Method of Resolution and Computer Program

The system of equations is solved in order to obtain all desired results. Considering the complexity of the system and the dependence between all variables, an iterative solving method is used. It always insures:

- a forces equilibrium:

$$\sum_{i=1}^{N2} Zpi(i,2) = FAXI$$

- a geometrical equilibrium $|\Delta LM| < \varepsilon$ where $\Delta LM$ is the difference between the length of the flat belt and the sum of strut thicknesses. Note that the strut thickness variations are calculated with the forces $X_i$ from a non linear experimental compression curve.

The calculation is done in three steps (Fig. 4):

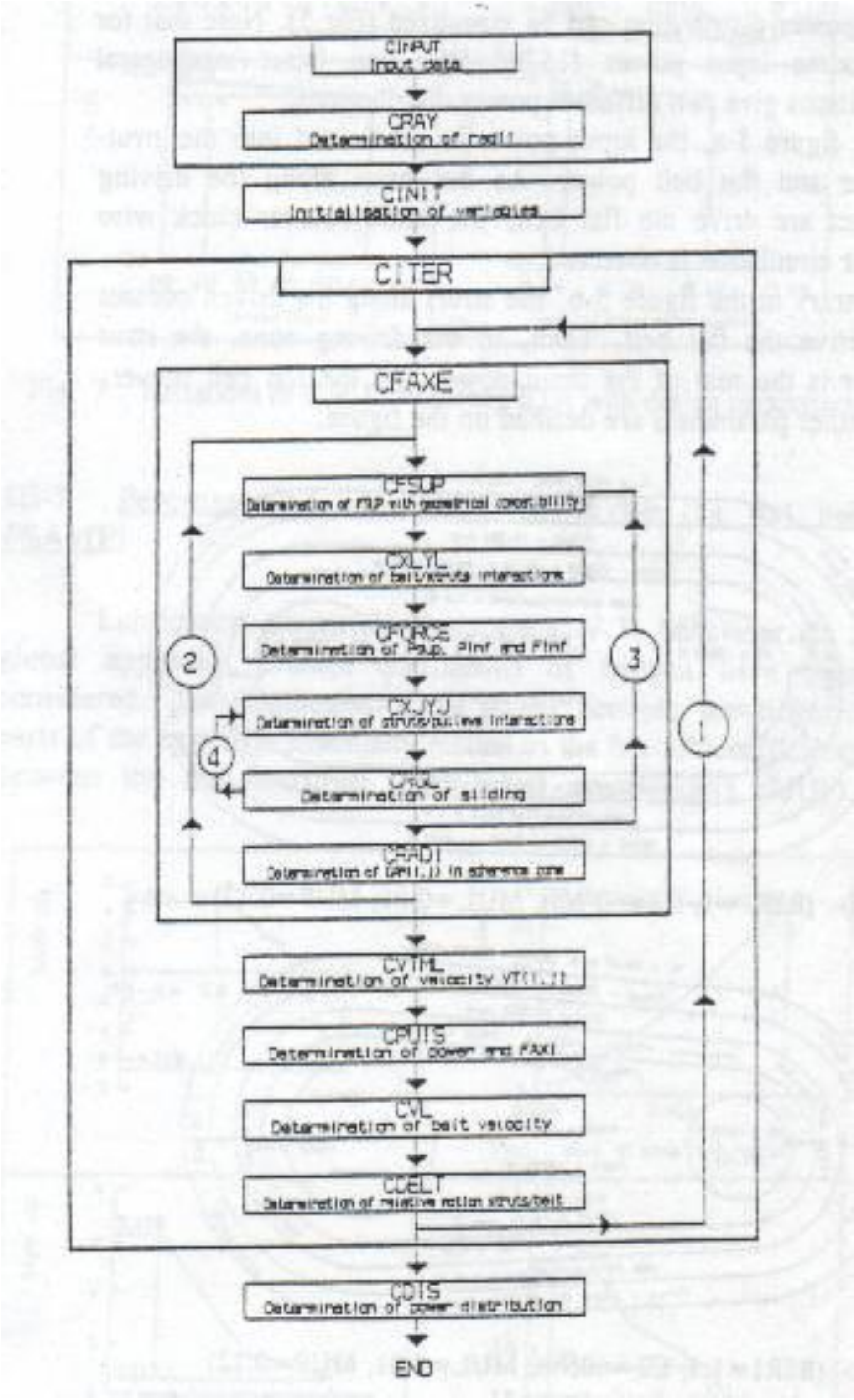

Fig. 4 : Flow chart of the computer softwere

- Input data by reading data files (INPUT),
- variables initialization considering a rigid model (CINIT),
- following variables calculations are made with iterative methods (CITER).

  The iterative process of the system equilibrium is done then into the computer module CITER. This module calculates:
- NG and NL by the equilibrium of torque (for NG) and the tension in the belt (for NL),
- FAXI, the force imposed on the driving pulley, and FAXE a force inter-axes pulleys (Loop 2). The numerical secant methode has been used in order to obtain the convergence between the sum of the calculated individual strut-pulley interactions $ZP_i$ and the imposed input value of FAXI,
- determination of FSUP which consists in turn of the determination of action XL, YL, PSUP, PINF, FINF, $X_i$, XP, YP, and the calculation of radial and tangential slidings of struts (that means $\gamma p_i$ angles).After the achievement of the geometric compatibility, the angle $\gamma$pi along the adherent zone are calculated from the computing module CRAD1. Finally, the velocity of struts and flat belt, the power distribution, the efficency, the relative motion of flat belt in relation to the struts and the distribution of powers are calculated.

  The power distribution can be visualized (Fig.5). Note that for the same input power (15706 W), two input mechanical conditions give two different power distributions:
- in the figure 5-a, the input power is subdivized into the strut-power and flat belt power. As the struts along the driving contact arc drive the flat belt, the same counter clock wise power circulation is observed,
- in contrary in the figure 5-b, the struts along the driven contact arc drive the flat belt. Then, in the driving zone, the strut power is the sum of the input power and the flat belt power.

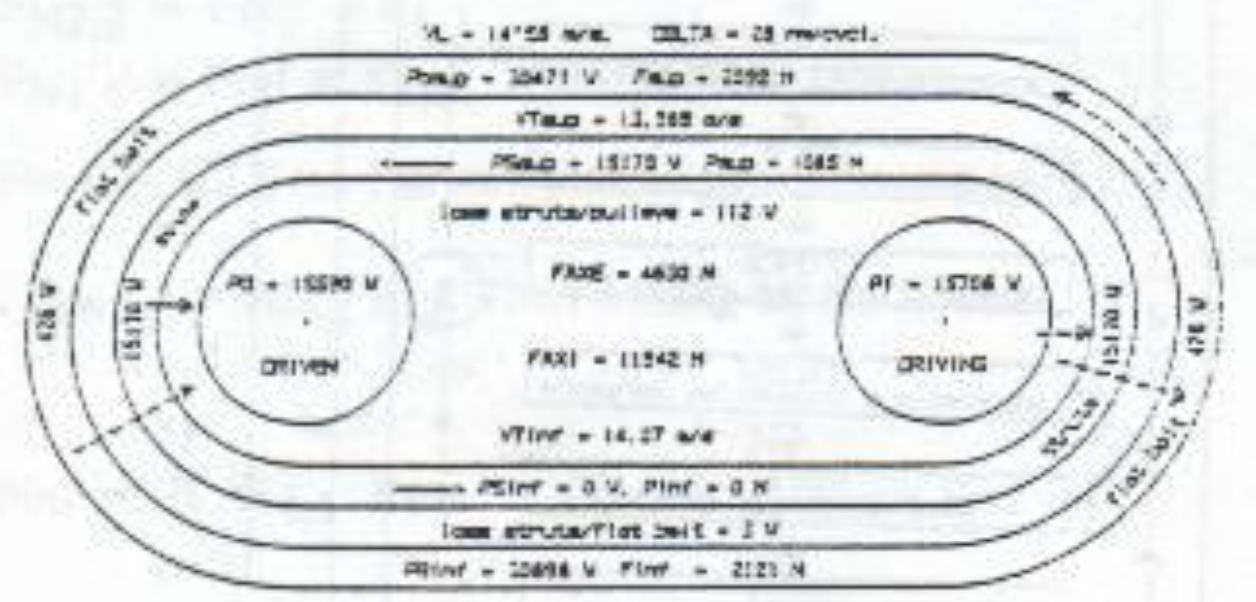

(a)- (R2R1=1, C2=60 Nm, MUL=0,01, MUP=0,12)

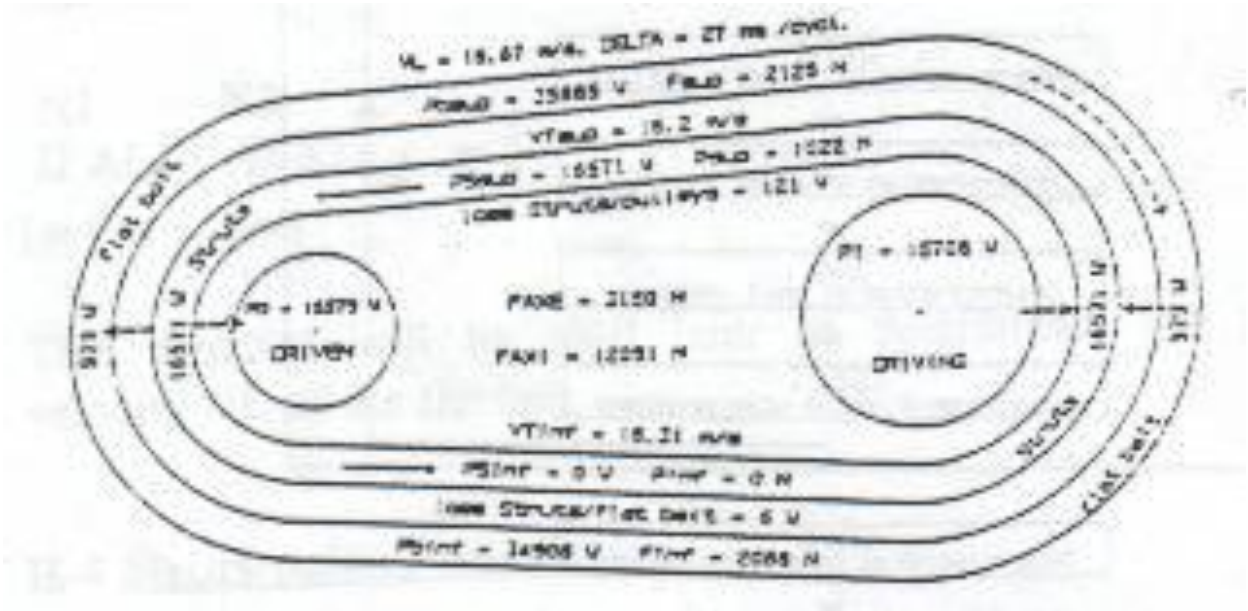

(b)- (R2R1=1,4, C2=60Nm, MUL=0,01, MUP=0,12)

Fig. 5 : The power distribution between different parts of system

## III - <u>Results and discussion</u>

For one reduction ratio, a fixed inter-axe, an input torque, an axial force imposed on the driving pulleys and an initial load F0 (An initial load is necessary to insure the initial compression of the struts), the computer program gives the following results (Fig. 6):

- tension in flat belt Fig. 6-a),
- normal action pulley/struts (NP Fig. 6-b),
- sliding angle ( $\gamma_p$ Fig. 6-c),
- velocity of struts (VT Fig. 6-d),
- pressure between struts (X Fig. 6-e),
- amplitude of the sliding displacement between struts and pulleys Y [ $Y = \sqrt{\text{GT2 + GR2}}$ ) ] (Fig. 6-f).
- tangential action flat belt/struts (XL Fig. 6-g),
- circumferential action pulleys/struts (XP Fig. 6-h),

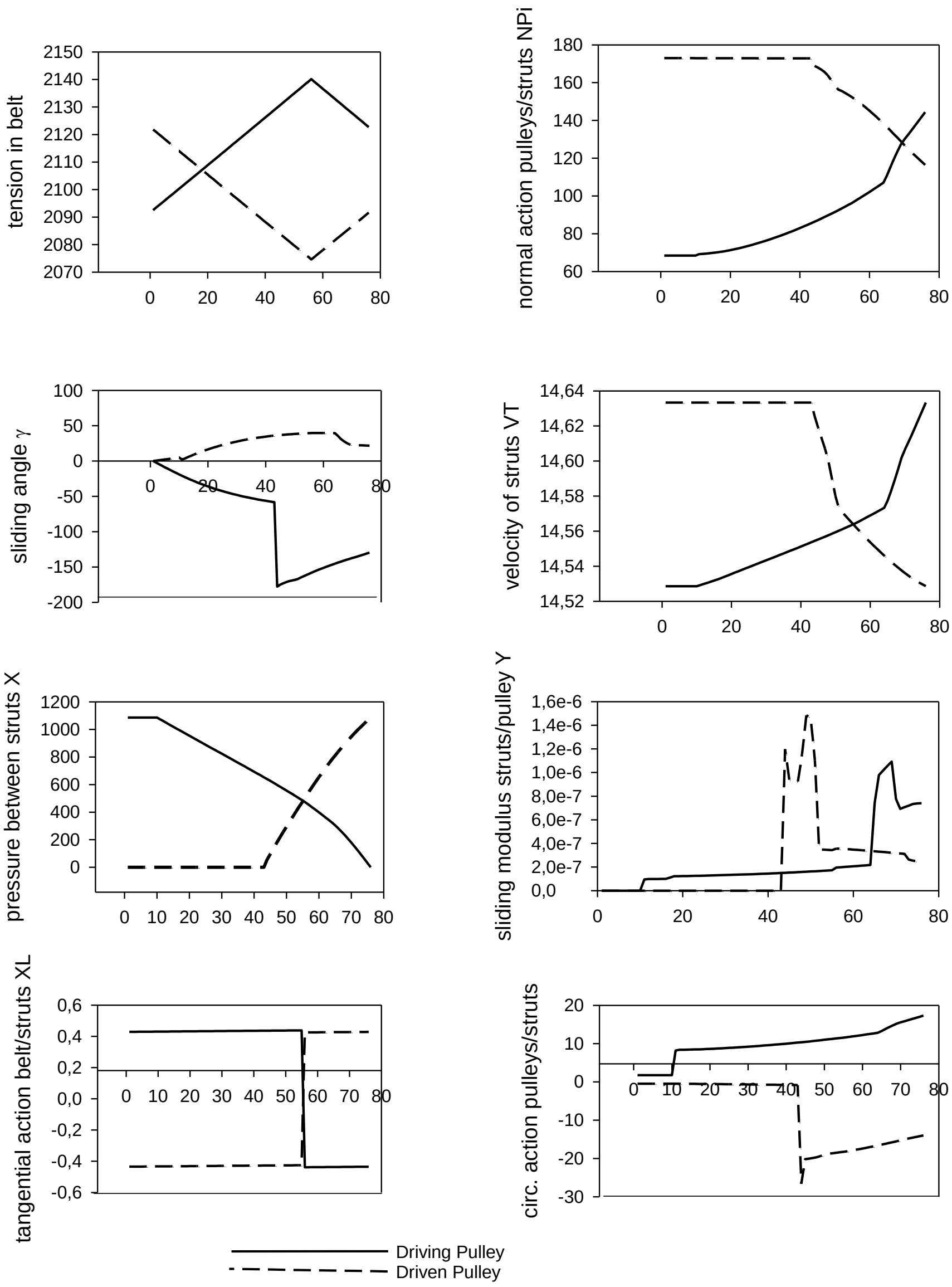

-

Fig. 6 : Numerical results issued from the computer, versus the position of struts along the contact arc (R2R1 = 1, C2 = 60 Nm, MUL = 0,01, MUP = 0,12)

Note that, on the figure 6:

- the pressure X between struts is constant along the adherence zone. It increases after that on the driving pulley (from the strut $NG_2$) and decreases on the driven pulley (from strut $NG_1$),
- a change of the sign of the tangential action (XL) when the relative motion flat belt /struts is reversed,
- the circonferencial action pulleys/struts (XP) depends on the sliding direction of struts in relation to pulleys. This action is positive on the driven pulley (the struts slides forward this pulley) and negative on the driving pulley (the struts slides backward),
- the tension in the flat belt increases until $NL_1$ and decreases after that to reach the value FSUP. On the driven pulley, this

evaluation is inverted on the driving pulley : decreasing until the strut $NL_2$, then increasing to arrived to the FINF value. Note that the ratio Finf/Fsup is very dependant of the coefficient of friction µL between belt/struts [7] while it is practically independant of the others parameters,

- the normal action pulley/struts (NP) is constant in the adherence zone (like the pressure between the struts). On the driving pulley, this action decreases, but it increases in the driven pulley. Therefore, this value stays more important on the driving pulley,
- the struts velocity VT varies in the same way as the normal action. In the adherence zones, velocity does not vary. In the figure, note that the velocity along the linear strand is not represented,
- the sliding angle (  p) on the driven pulley is comprised between 0° and 90°, the strut is backward pulley. On the driving pulley, it is in the sliding zone between -90° and -180°, the strut is then forward,
  - amplitude of the sliding displacement of a strut in relation to VT and   p increases on the driven pulley (in the sliding zone) and decreases in the driving pulley.

The computer program gives also other results as the efficency of the mechanical system, the percentage of the transmitted power by the flat belt and the struts, the way of distribution of power in the different parts of the system, the relative motion between the flat belt and the struts.

## III-1 Influence of input parameters

The input parameters such as speed or reduction ratio R2R1, input torque C2, the friction coefficients MUL and MUP influence the results. In order to know the variation of the following results:

- efficency,
- percent of power transmitted by the flat belt,
- tension in the lower strand of the flat belt,
- pressure between the struts in the upper strand.
- axial force imposed on the flanges of driving pulley,

systematic numerical tests have been made in order to visualize tendencies depending on these parameters. When one parameter varies the others are kept constant and equal to : R2R1=1, C2=80Nm, MUL=0,01, MUP=0,12.

## III-2 The Efficiency

In all cases, the general efficiency is high (more than 97% approximately). It is due to the hypothesis on power loss where different causes are ignored (contribution of friction between struts, hysteresis of materials, polygonal effect). Efficiency varies a little with the input parameters. In fact, for a reduction ratio R2R1 smaller than 1, the efficiency increases, but for a mulptiplicator C.V.T., the efficiency decreases with the ratio (Fig.7). This result is in accordance with experimental results and with conclusions of the litterature. For a ratio equal 1, the efficiency is pratically constant versus the torque (fig.7) and with the struts/pulley friction coefficient (fig.7) and increases with the flat belt/struts coefficient of friction.

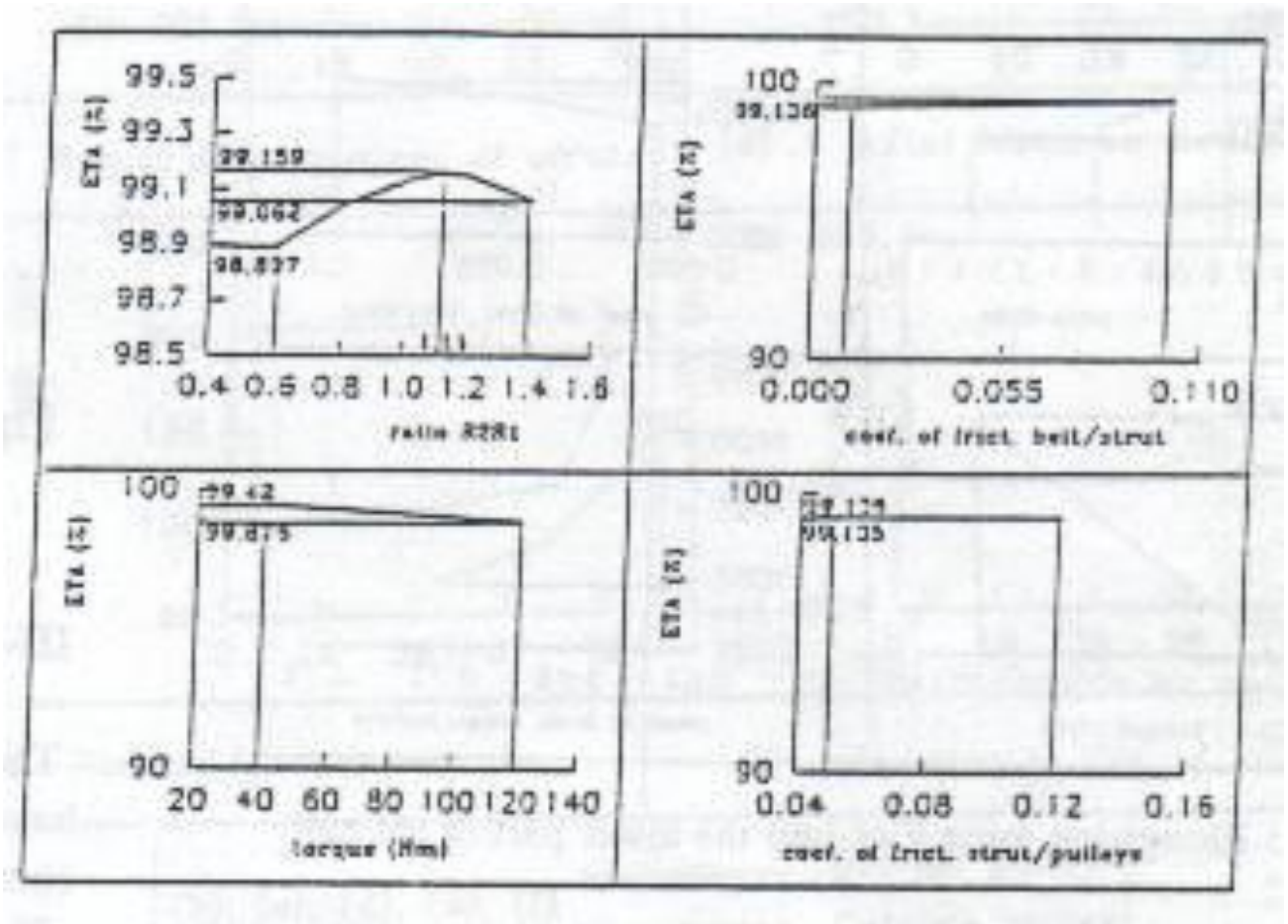

Fig. 7 : Variations of C.V.T. efficiency ETA with design parameters

## III-3 Percentage of transmitted power by the flat belt (%AME)

Lubrication govern obviously the C.V.T. behaviour. In a global approach, average coefficients of friction have been considered. The distribution of the power between the different parts of the system is essentially related to the friction coefficients between the flat belt/struts (MUL) and struts/pulleys (MUP). Considering the friction coefficient between the flat belt and the struts, the percentage of transmitted power by the flat belt increases in relation to the coefficient (MUL). The parameter %AME decreases for a ratio R2R1 which increases between 0,8 and 1,4 and for a coefficient of friction struts/pulley between 0,05 and 0,12 (fig. 8).

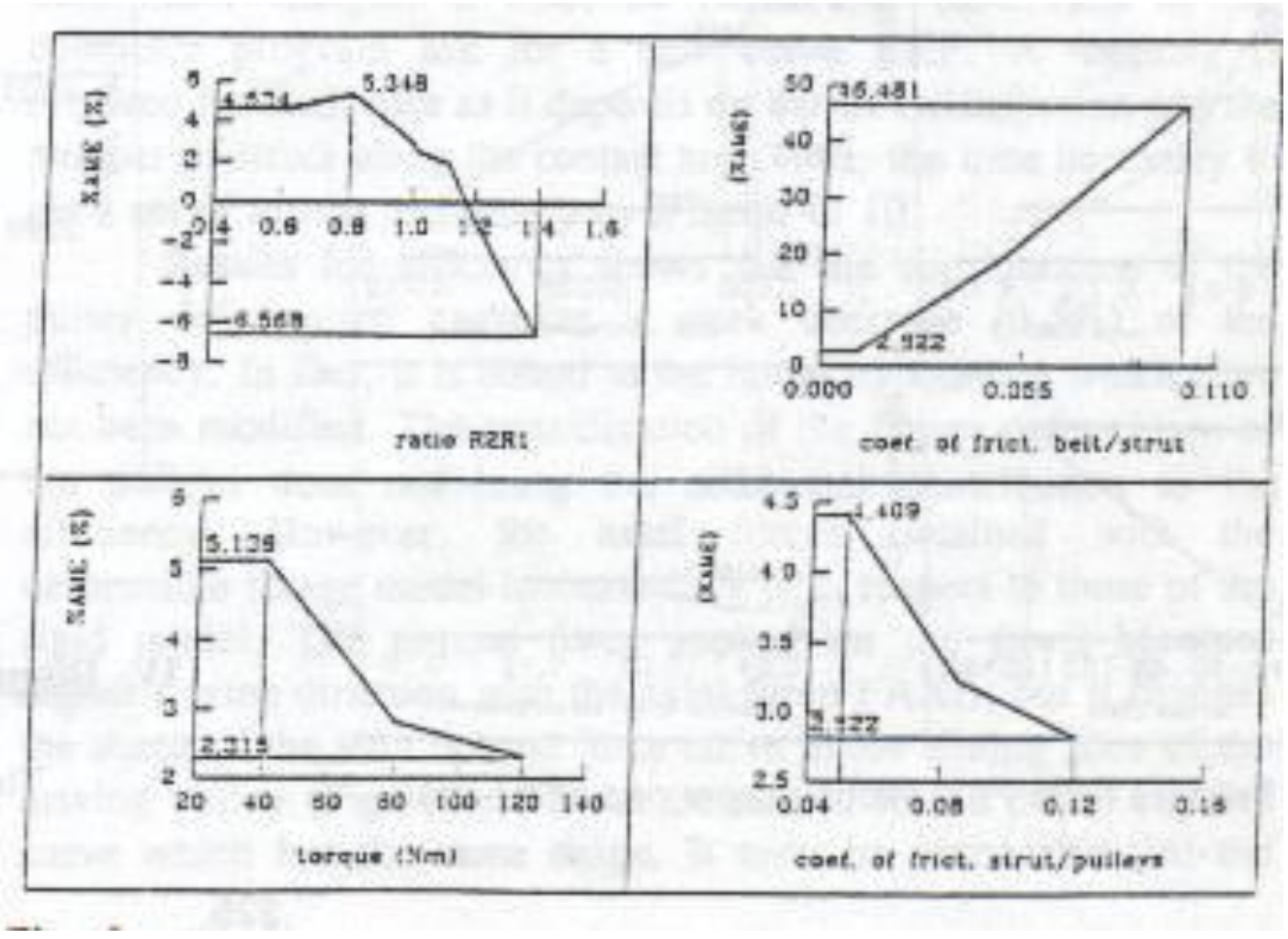

Fig. 8 : Variations of the inner power %AME transmitted by the flat belt

### III-4 <u>The tension in the lower strand of the flat belt (FINF)</u>

The input torque is the more influent parameter on FINF. This tension FINF increases with this torque $C_2$ (Fig. 9) as the percentage of power was shown to increase with torque. It decreases with the R2R1 ratio and the increasing of the struts/pulleys and flat belt/struts friction coefficients.

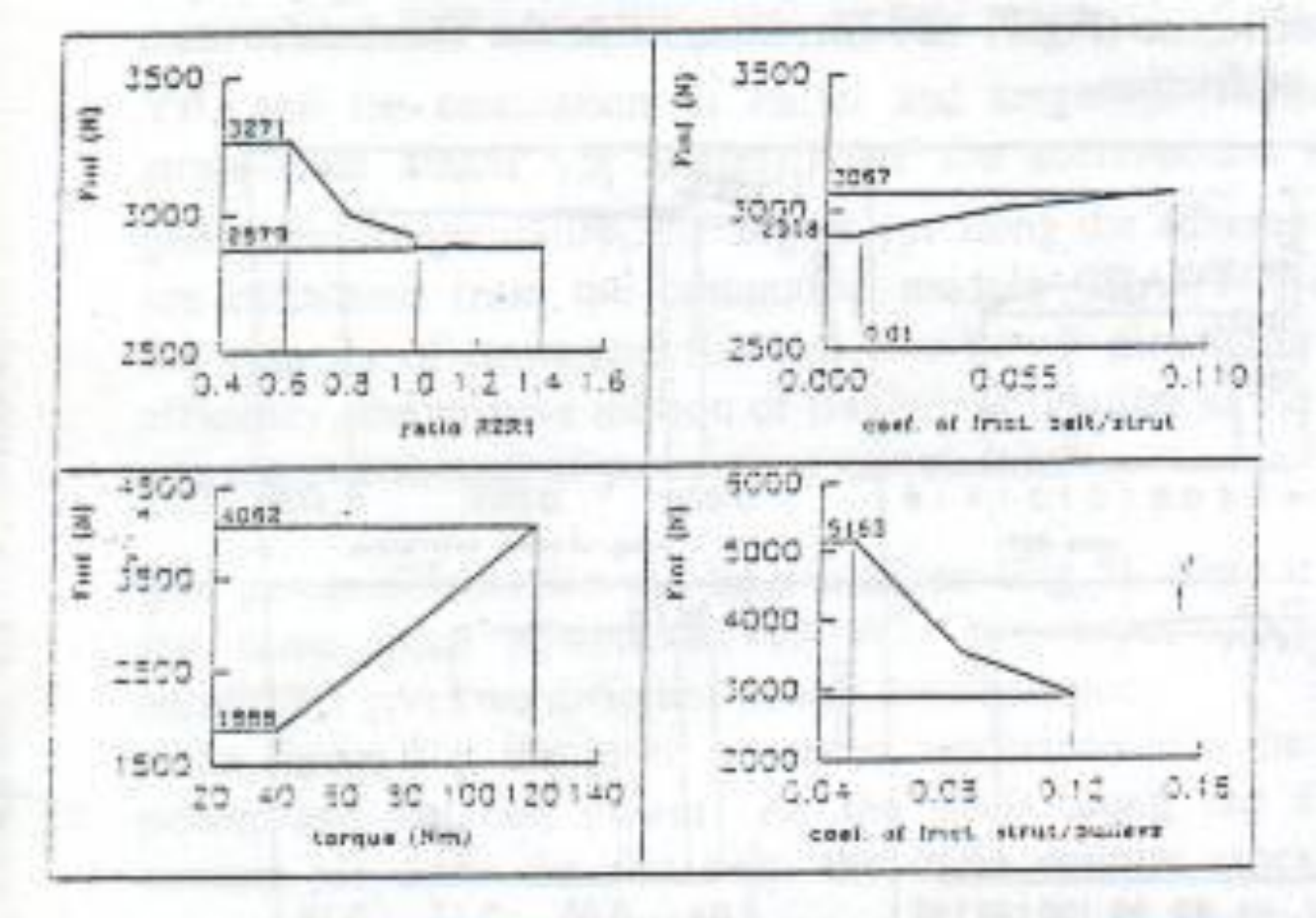

Fig. 9 : Elongation force Finf into the lower part of the belt

### III-5 <u>The pressure between the struts in the upper strand (PSUP)</u>

The pressure between the more compressed struts (in the upper strand) decreases with an increasing of reduction ratio (Fig. 10) and flat belt/struts friction coefficient (Fig. 10). This pressure increases with the input torque (Fig. 10) and the struts/pulleys friction coefficient(Fig. 10).

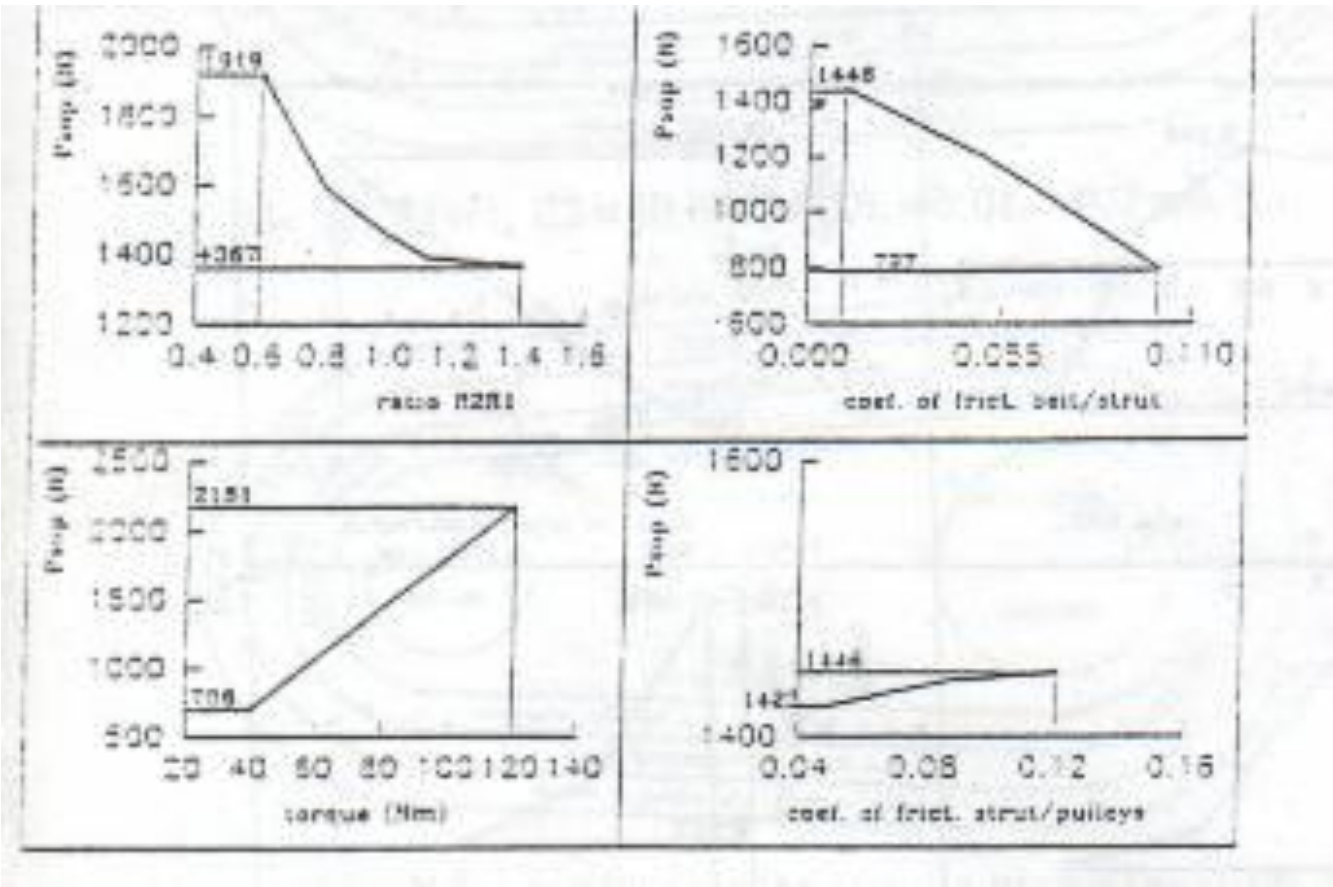

Fig. 10 : Elongation fore Psup into the upper part of the struts

## III-6 <u>The axial force imposed on the flanges of driving pulley (FAXI)</u>

Remember that the first part of this paper considers rigid flanges of the pulleys. The force applied on the flanges of the pulleys FAXI is governed by the input parameters (Fig. 11). FAXI decreases when the ratio R2R1 increases. The same trend is observed with the variations of the friction coefficients. FAXI increases obviously with torque.

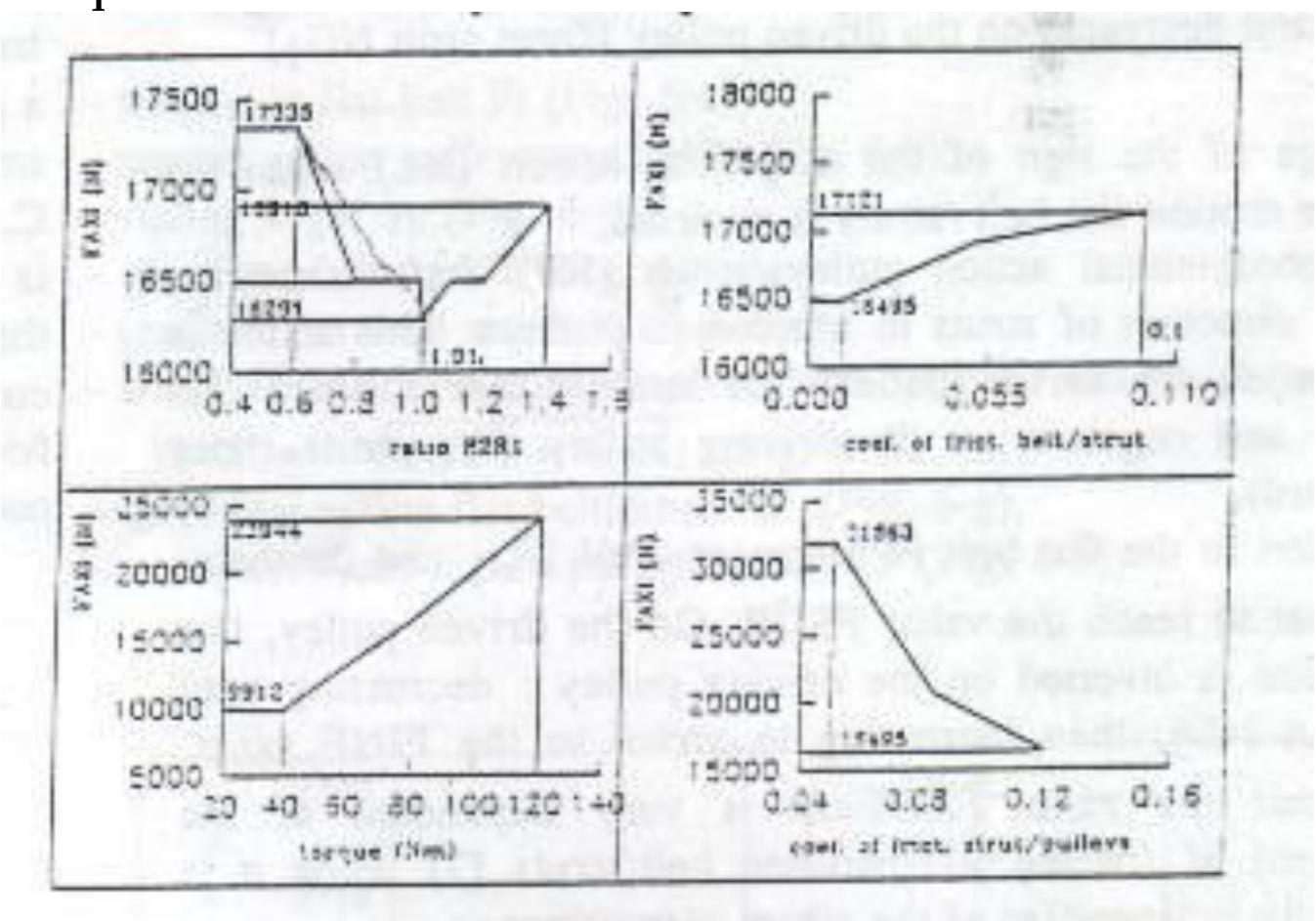

Fig. 11 : Variation of C.V.T. axial force FAXI with design parameters

## III-7 <u>The initial tension F0</u>

The initial tension F0 is acheived by adding struts when the belt have a global circular shape before mounting on the pulleys. Two zones are isolated from the figure, the larger F0 gives the higher efficiencies. The other parameters increases with F0 (Fig. 12). From a global point of view, the non linear deformations of the struts ask for a sufficient initial F0 in order to limit compressive displacement variations. It must be also pointed out that the flat belt becomes influent only for large F0.

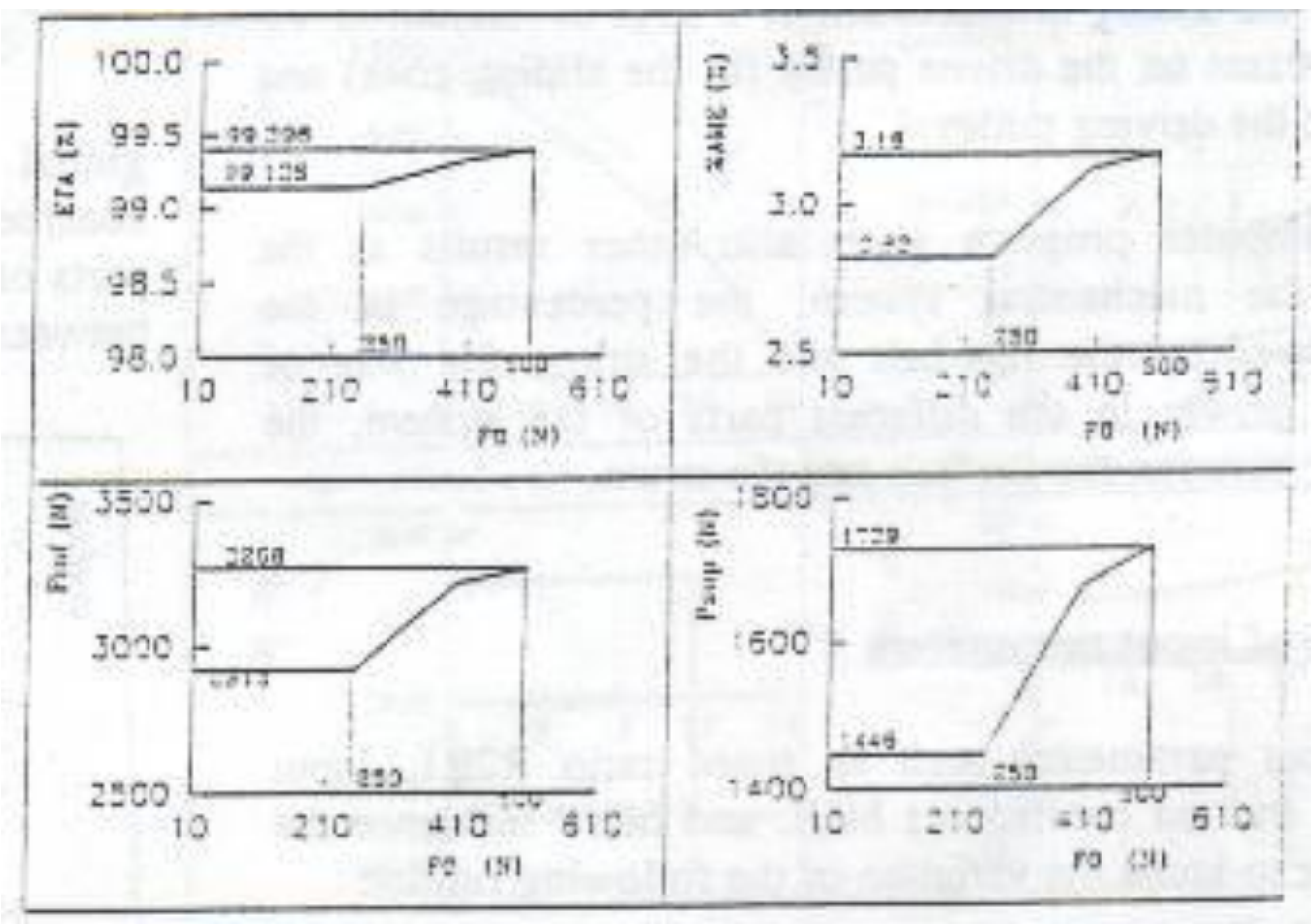

Fig. 12 : Initial tension F0 influence

## IV- Discussion

Incerted in the contacting arc, the strut (i) exerts an axial force $ZP_i$ on the flange of the pulley. This forces provokes a deformation, and concequently, the flange displaces axially of the quantity $EKF(ZP_i)$. so, the flange has gaps with respect to the position (i-1) of the quantity:
$EKF(ZP_i) - EKF(ZP_{i-1})$ where EKF is the fonction of the flange displacement by terms of the applied force of the pulley and the considered strut. This function has been determinated by the Finite Element program (AMME) [8,9,10]). It is noted that this function is not lineare. The deformation of the flange leads an additional sliding of the strut in the pulley grouve. In effect, the position (i-1) to another one (i), the flanges of the pulley have gaps of the quantity $\Delta ai$ ($\Delta ai = 2[EKF(ZP_i) - EKF(ZP_{i-1})]$). It provokes a radial sliding of the strut with a quantity GRPi:
$GRPi = - [\Delta ai]/[2\sin(\phi)]$. The deformation of the struts provokes the radial sliding at the quantity GRMi:
$GRMi = -[EKL(ZPi)-EKL(ZPi-1)]/[2\sin(\phi)]$. So the radial sliding GRi becomes the som: $GRi = GRPi + GRMi$.

In the adherent zone, we supposed that the force ZPi rests constants because the displacements induced by the flange deformations are small. Consequently, a radial sliding curve of the struts (due to the pulley deformation) is obtained.

## V- Conclusion

The metal V-belt is a highly discretized mechanism with a large number of parts. The relative motion of the elements adds to the difficulties of the mechanical behaviour modelization.

A theoretical study has been made with the equations of the general mechanics in a quasi-static working case. The originality of this approach lies on the consideration of the whole system constituted by an large number of interactive components. A computer software has been developped in order to make numerical simulations as the cost of experimental simulations become too expensive. The results have shown that:

- the distribution of power between the different parts of the system is governed by the control of the friction coefficients,
- the efficiency of the system depends essentially on the rigidity of the materials of struts. The use of less rigid materials than the steel struts reduce the efficiency,
-the initial tension F0 has an influence on the results from a lower limit.

**Acknowledgement**

A part of this work have been made under industrial contract and the authors wish to thank Mr Douhairet and Trinquard for their helpfull comments.